\documentclass[%
 reprint,
 amsmath,amssymb,
 aps,
]{revtex4-1}

\usepackage{graphicx}
\usepackage{dcolumn}
\usepackage{bm}

\begin{document}

\bibliographystyle{apsrev4-1}

\title{Surface-acoustic-wave-controlled optomechanically induced transparency in a hybrid piezo-optomechanical planar distributed Bragg-reflector-cavity system}
\author{Shi-Chao Wu}
\affiliation{Shanghai Advanced Research Institute, Chinese Academy of Sciences, Shanghai, 201210, China}
\affiliation{University of Chinese Academy of Sciences, Beijing, 100000,China}
\affiliation{Jiangsu Ocean University, School of Sciences, Lianyungang, 320700, China}
\author{Li Zhang}
\affiliation{Shanghai Advanced Research Institute, Chinese Academy of Sciences, Shanghai, 201210, China}
\affiliation{University of Chinese Academy of Sciences, Beijing, 100000, China}
\author{Jian Lu}
\affiliation{Shanghai Advanced Research Institute, Chinese Academy of Sciences, Shanghai, 201210, China}
\author{Li-Guo Qin}
\affiliation{Shanghai Advanced Research Institute, Chinese Academy of Sciences, Shanghai, 201210, China}
\affiliation{Department of Physics, Shanghai University Of Engineering Science, Shanghai, 201620, China}
\author{Zhong-Yang Wang}
\email{wangzy@sari.ac.cn}
\affiliation{Shanghai Advanced Research Institute, Chinese Academy of Sciences, Shanghai, 201210, China}

\begin{abstract}
 We propose a scheme that can generate tunable optomechanical induced transparency in a hybrid piezo-optomechanical cavity system. The system is composed of a high-quality planar distributed Bragg reflector cavity modified with an embedded Gaussian-shaped defect. Moreover, interdigitated transducers are fabricated on the surface of the cavity to generate surface acoustic waves. Under the actuation of the surface acoustic wave, the upper Bragg mirrors can be vibrated as a bulk acoustic resonator, and the distributed Bragg reflector cavity becomes a standard three-level optomechanical system. In this situation, we show that when a strong pump optical field and a weak probe optical field are simultaneously applied to the hybrid optomechanical cavity system, optomechanically induced transparency occurs under the quantum interference between different energy level pathways. Our scheme can be applied in the fields of optical switches and quantum information processing in solid-state quantum systems.
 
\end{abstract}


\maketitle

\section{Introduction}

In the field of quantum information, many types of artificial atomic systems have been proposed to regulate and manipulate the processing of optical or microwave information, including superconducting quantum circuits \cite{RMPHybridquantumXiang,PRLSuperconductingCleland}, N-V centers \cite{2014prlNVcenter,NVCenterprl2016}, semiconductor quantum dots \cite{UltrabrightsourceSemiqudot2010,2012prlquantumdot}, and distributed Bragg reflector (DBR) cavities \cite{2011prldbr,2013NCDBR}. Furthermore, since most of these artificial atomic systems can be coupled with both photon and phonon modes, the optical information can also be operated by modulating the phonon modes. There are many methods to control the phonon modes; for example, phonon modes can be driven by the radiation pressure of the optical field or microwave field \cite{andrewsNaturebidirectionalconver,PRAEntanglingBarzanjeh}. Moreover, in recent experiments, the phonon modes can be driven by different types of external mechanical fields, such as the Lorentz force and piezoelectric force \cite{Alexei2005Evidence,naturenanomechanicalBochmann}. A representative mechanical field is the surface acoustic wave (SAW) \cite{PhononinduceOpticalsuperlatticeprl2005,PhysRevLett.105.037401,SAWreviewLima2005,AcoustoOpticOptomeCirPrapplied2017,
doi:10.1021/nl1042775,prb2001saw,SAWprb2001,UniversalQuantumTransducersPhysRevX2015,doi:10.1063/1.4908248}, which is generated by the piezoelectric effect. Since piezoelectric materials are widely used in optically DBR cavity systems, SAW can be applied to modulate the optical properties of DBR cavity systems by controlling the phonon modes \cite{PhysRevLett.105.037401}.

The SAW is a type of phonon-like excitation containing mechanical vibration modes \cite{PhononinduceOpticalsuperlatticeprl2005,PhysRevLett.105.037401,SAWreviewLima2005,AcoustoOpticOptomeCirPrapplied2017,
doi:10.1021/nl1042775,prb2001saw}. It is electrically generated on the surface of the piezoelectric substrate using interdigital transducers (IDTs), which are driven by the radio frequency voltage source (RF). Furthermore, SAW contains both transverse and longitudinal vibration modes; when it propagates along the surface of the substrate, it simultaneously propagates downward. The velocity of the SAW propagating along the surface is determined by the material performance of the substrate. When the SAW propagates downward, its vibration frequency can be adjusted with the design of the IDTs, and the extension depth below the surface is approximately one wavelength of the SAW. Accordingly, the surface of the substrate in the vertical direction is vibrated as a bulk acoustic resonator (BAR), which can be considered a standard mechanical resonator \cite{NVCenterprl2016}. Moreover, the intrinsic damping rate of the BAR is determined by the SAW mode, which is equal to the line width of the SAW mode. In the related research, the quality factor of the BAR can exhibit the order of $10^5$. Correspondingly, its frequency can exceed the order of GHz, and its intrinsic damping rate can exceed the order of 10 kHz \cite{SAWAPL2003,UniversalQuantumTransducersPhysRevX2015,doi:10.1063/1.4908248}.

Planar DBR cavity systems have been applied to realize light-sound interactions in related experiments \cite{lightsoundinteractionapl2004}. Generally, Generally, the quality of the DBR cavities can be improved by increasing the number of mirror pairs or optimizing the design of the structure \cite{apl2004defect,apldefect2009,Verticalprb2013}, and a representative type of cavity structures is the DBR cavity structure modified with the embedded sub-micrometer Gaussian-shaped defect. This structure is proposed by F. Ding et al.; relative to traditional DBR structures with the same number of mirror pairs, its quality factor can be effectively improved by nearly 2 orders of magnitude \cite{Verticalprb2013}. Physically, when the Gaussian-shaped defect is sandwiched between the upper and the lower Bragg mirrors, due to the local change in cavity length and the deformation of the photonic defect band, the photons are confined in a small modal volume in the vertical direction. Moreover, the light scattering induced by the lateral dielectric discontinuities is minimized. As a result, the quality of this structure dramatically improves. Based on the finite difference time domain (FDTD) calculation, the quality factors $Q$ of the designed DRR cavity can exceed the order of $10^5$ \cite{Verticalprb2013}.

Here, we propose a hybrid planar DBR cavity system, where IDTs are fabricated on the surface to generate the SAW. Moreover, the quality factor of the DBR cavity is optimized by the embedded Gaussian-shaped defect, and the DBR cavity is composed of ${\rm {AlAs}}$ and ${\rm {GaAs}}$ alternating layers, which are piezoelectric materials to support the SAW generation. When the SAW is applied to the system, we can estimate that the thickness of the upper Bragg mirrors is within the wavelength scope of the SAW. As a result, the upper Bragg mirrors in the vertical direction vibrate as a BAR, which can be considered a mechanical resonator. Accordingly, our system is turned into a standard three-level optomechanical cavity system \cite{RevMooptomechanicsAspelmeyer,OptomechanicallyInducedWeis2010}, which is formed by the energy levels of the DBR cavity and BAR.

The traditional three-level optomechanical cavity system is composed of an optical cavity and a mechanical resonator; it has been applied in many fields, and a representative system is the optomechanically induced transparency (OMIT) \cite{Fan2015Cascaded,Bienert2015Optomechanical,Stannigel2011Optomechanical,Xu2015Mechanical,Si2017Optomechanically,Ma2015Optomechanically,
2015tunablefast,Wang2015Precision,PRA2014twocolorWang,PRAcoupledresonatorsDuan,PRATunableMa}. OMIT is a special optical phenomenon; when it occurs, the susceptibility of the optomechanical cavity system changes; then, the resonance optical field can be modulated from opacity to transparency. Physically, it arises from the quantum interference among different energy-level pathways \cite{PRAEITAgarwal,NatCoherentBalram,safavinatureelectromagnetically,PRAPhaseJia}. Similar to the electromagnetically induced transparency (EIT) observed in three-level atomic systems \cite{PhysicsTodayHarris,PRLEITBoller}, OMIT has been applied in many fields including quantum information processing \cite{Stannigel2011Optomechanical}, fast and slow light \cite{2015tunablefast} and quantum optical storage \cite{Mcgee2013Mechanical}.

In our system, we show that when a strong pump optical field and a weak probe optical field are simultaneously applied to the hybrid optomechanical cavity system, in the presence of the SAW, a transmission window can be obtained in the weak output probe field. This phenomenon arises because under the driving of the SAW, the upper Bragg mirrors vibrate as a BAR; then, the DBR cavity becomes a lambda-type three-level optomechanical system. Under quantum interference among different energy level pathways, OMIT occurs; as a result, the transmission window is observed in the weak output probe field. Conversely, without the actuation of the SAW, the transmission window disappears. 

Compared to traditional micro-nano optomechanical cavity systems, we propose an artificial DBR optomechanical cavity system that can be applied in optical quantum information processing, such as optical switches and information storage. Moreover, our system has many advantages: (i) The system parameter settings of our system are flexible; in particular, both frequencies of the optical cavity and nanomechanical resonator can be autonomously and accurately designed according to the application requirements. (ii) Our system combines radio-frequency and optical signals to realize the electrically controlled quantum optical information processing. (iii) The relevant manufacturing technology of the DBR microcavity our system adopted is mature and feasible, and our system can be combined with quantum dots and quantum traps to expand to the field of quantum device integration. 

The paper is organized as follows. In Sec. II, we describe the proposed model and derive the system Hamiltonian. In Sec. III, we present the dynamical process of the system. In Sec. IV, we discuss the detailed physical mechanism of the OMIT and study the change in output field controlled by the SAW. Finally, we make a brief conclusion in Sec. V.

\section{MODEL AND HAMILTONIAN OF THE SYSTEM}

\begin{figure}
\centering
\includegraphics[angle=0,width=9cm]{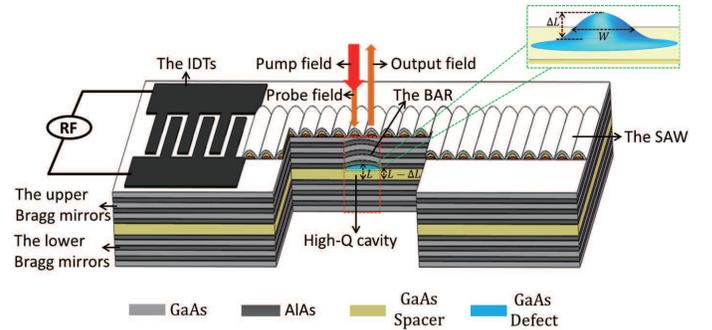}
\caption{(color online) Schematic diagram of the hybrid piezo-optomechanical DBR cavity system, where the IDTs are fabricated on the surface of the cavity to generate the SAW and the IDTs are driven by the RF. The high-Q planar DBR cavity (the area inside the red dashed box) is modified with the embedded Gaussian-shaped defect. The cavity is composed of ${\rm {AlAs}}$ (dark-grey color area) and ${\rm {GaAs}}$ (light-white color area) alternating layers, and the ${\rm {GaAs}}$ spacer (dark-yellow color area) is sandwiched between the Bragg mirrors. $L$ denotes the thickness of the DBR cavity, and $\triangle L$ and $W$ denote the thickness and half-width of the Gaussian-shaped defect (blue area), respectively. The optomechanical cavity is driven by a strong optical pump field (red arrow) and a weak optical probe field (brown arrow).}
\end{figure}

\begin{figure}
\centering
\includegraphics[angle=0,width=9cm]{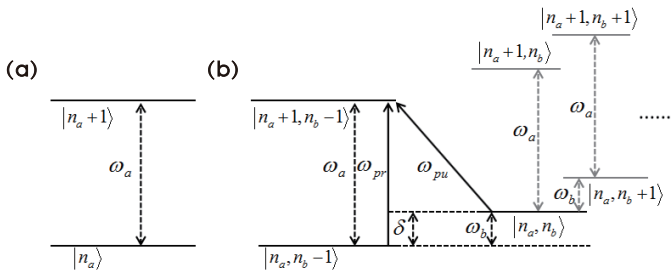}
\caption{(a) Energy level structure of the hybrid optomechanical system (a) without and (b) with the SAW mode, where $|n_a\rangle$ and $|n_b\rangle$ are the number states of photons and phonons, respectively. The energy difference between $|{n_a}\rangle$ and $|{n_a+1}\rangle$ is the frequency of the DBR cavity $\omega_a$, and the energy difference between $|{n_b}\rangle$ and $|{n_b+1}\rangle$ is the frequency of the BAR $\omega_b$. Here, $\omega_{pr}$ is equal to $\omega_{a}$, with the relation $\omega_{pr}-\omega_{pu}=\delta=\omega_{b}$.}
\end{figure}

The hybrid piezo-optomechanical planar DBR cavity system that we propose is illustrated in Fig. 1, where IDTs are fabricated on the surface of the cavity to generate the SAW. The IDTs are driven by the RF, which can convert the RF voltage signal to SAW via the piezoelectric effect. The high-Q cavity planar DBR cavity is modified with the embedded Gaussian-shaped defect, as designed by F. Ding et al. to improve the quality factor of the cavity \cite{Verticalprb2013}. The optomechanical cavity is driven by a strong optical pump field and a weak optical probe field. In detail, in our system, the quarter-wavelength design for the DBR mirrors is used to maximize the photon confinement. The DBR cavity is composed of ${\rm {AlAs}}$ and ${\rm {GaAs}}$ alternating layers. The ${\rm {GaAs}}$ spacer is sandwiched between the upper (10 pairs) and the lower (15 pairs) Bragg mirrors, and the same structure is applied in the recent experiment \cite{PhysRevLett.105.037401}. 

Correspondingly, based on the design of the embedded Gaussian-shaped defect DBR cavity structure, the photons on resonance are confined in a small modal volume on the order of $(\lambda/n)^3$ ($\lambda$ is the spacer optical thickness, and $n$ is the refractive index of the spacer material), and the light scattering induced by the lateral dielectric discontinuities is minimized. As a result, the quality factor is improved by nearly 2 orders of magnitude relative to the traditional DBR cavity structure \cite{Verticalprb2013}. 

First, we consider the situation where only a pump optical field and a probe optical field are simultaneously applied to the DBR cavity, and the frequencies of optical fields are referred to as $\omega_{pu}$ and $\omega_{pr}$, respectively. Moreover, we assume that the phase difference between the probe and pump fields is zero, and the phase difference is locked. In this situation, the DBR cavity is a standard two-level energy system, whose Hamiltonian is given by
\begin{align}
&H_I={\omega}_{a}{\hat a}^{\dag}{\hat a}+  \nonumber \\
&{i}{\hbar}{\varepsilon}_{pu}(\hat{a}^{\dag}e^{{-i}\omega_{pu}t}-\hat{a}e^{{i}\omega_{pu}t})+{i}{\hbar}{\varepsilon}_{pr}(\hat{a}^{\dag}e^{{-i}\omega_{pr}t}-\hat{a}e^{{i}\omega_{pr}t}).
\end{align}
Here, the first term of $H_I$ describes the energy of the cavity mode, where $\omega_{a}$ is the frequency of the DBR cavity mode. ${\hat a}^\dag$ and ${\hat a}$ are the creation and annihilation operators of the cavity mode, respectively. The second and third terms describe the energies of the input optical fields, and $\varepsilon_{pu}={\sqrt{P_{pu}\kappa_{a}/(\hbar\omega_{pu})}}$ and $\varepsilon_{pr}={\sqrt{P_{pr}\kappa_{a}/(\hbar\omega_{pr})}}$ are the amplitudes of the optical pump field and probe field, respectively. $P_{pu}$ and $P_{pr}$ are the input powers of the optical pump field and optical probe field, respectively. $\kappa_a$ is the decay rate of the optical DBR cavity.

When the RF is also applied to the IDTs, the SAW is generated, and the upper Bragg mirrors vibrate as a BAR. Then, the Hamiltonian of the BAR can be written as 
\begin{flushleft}
\begin{eqnarray}
\begin{split}
H_b=\frac{{\hat p}^2}{2m_b}+{\frac{1}{2}}m_b{\omega_b}^2{\hat q}^2,
\end{split}
\end{eqnarray}
\end{flushleft}
where ${\hat q}$ and ${\hat p}$ are the position and momentum operators of the BAR, $m_b$ is the mass of the BAR, and $\omega_b$ is the frequency of the BAR, which is equal to the frequency of RF. When the upper Bragg mirrors are vibrated as a BAR, the thickness of the DBR cavity $L$ is changed accordingly. 

In our system, the driving power of the RF transmitted by the IDTs is represented as $P_{rf}$.Referring to the relevant study \cite{NVCenterprl2016}, we also assume that the SAW energy is uniformly distributed between the surface and one wavelength deep; then, the cross-sectional area through which the SAW travels is determined as $A=l_{\rm{IDTs}}\lambda_{s}=l_{\rm{IDTs}}\frac{v_{\rm{SAW}}}{\omega_b}2\pi$, where $l_{\rm{IDTs}}$ is the length of the IDTs fingers, $v_{\rm{SAW}}$ is the velocity of the SAW. According to the relevant research \cite{NVCenterprl2016}, the vibration amplitude of the BAR is given as
\begin{eqnarray}
\begin{split}
{q_{0}}=\sqrt{\frac{P_{rf}}{4 \pi l_{\rm{IDTs}} v_{\rm{SAW}}^2 \rho_{b} \omega_b}},
\end{split}
\end{eqnarray}
where $\rho_{b}$ is the average density of the BAR. Based on a recent experiment \cite{PhysRevLett.105.037401}, the amplitude of the BAR in our system can exceed 1 pm.

For the BAR whose vibration amplitude is $q_0$, referring to the relevant study \cite{NVCenterprl2016}, the total energy of it is given as $E_{b}=2m_b{{\omega}_b}^2{q_0}^2$. Taking the derivative of energy $E_{b}$ with respect to position, we can obtain the amplitude of driving force induced by the RF, which is
\begin{eqnarray}
\begin{split}
F_{rf}=4m_b{{\omega}_b}^2{q_0}.
\end{split}
\end{eqnarray}
Since the frequency of RF is also $\omega_b$, according to the relevant research \cite{NVCenterprl2016}, the Hamiltonian of the system can be written as
\begin{align}
&H_O=(1-{\frac{\hat q}{L}}){\omega}_{a}{\hat a}^{\dag}{\hat a}+
{i}{\hbar}{\varepsilon}_{pu}(\hat{a}^{\dag}e^{{-i}\omega_{pu}t}-\hat{a}e^{{i}\omega_{pu}t})\nonumber \\ &+{i}{\hbar}{\varepsilon}_{pr}(\hat{a}^{\dag}e^{{-i}\omega_{pr}t}-\hat{a}e^{{i}\omega_{pr}t})
+\frac{{\hat p}^2}{2m_b}+{\frac{1}{2}}m_b{\omega_b}^2{\hat q}^2\nonumber\\
&-F_{rf}q_{0}sin(\omega_{b}t).
\end{align}
Here, we assume that the phase difference between RF and optical fields is zero, and the phase difference is locked.

Now, we introduce the creation and annihilation operators for the BAR, which are
\begin{eqnarray}
\begin{split}
{\hat q}=\sqrt{\frac{\hbar}{2\omega_b m_b}}({\hat b}+{\hat b}^\dag),\\
{\hat p}=i\sqrt{\frac{\hbar \omega_b m_b}{2}}({\hat b}-{\hat b}^\dag),
\end{split}
\end{eqnarray}
where ${\hat b}^\dag$ and ${\hat b}$ are the creation and annihilation operators of the BAR, respectively. Then the Hamiltonian is rewritten as
\begin{align}
&H_O=\hbar{\omega}_{a}{\hat{a}}^{\dag}{\hat a}+\hbar \omega_b{\hat{b}}^{\dag}{\hat b}-\hbar{g_{om}}\hat{a}^{\dag}\hat{a}({\hat b}+{\hat b}^\dag) \nonumber\nonumber \\
&+{i}{\hbar}{\varepsilon}_{pu}(\hat{a}^{\dag}e^{{-i}\omega_{pu}t}-\hat{a}e^{{i}\omega_{pu}t})\nonumber +{i}{\hbar}{\varepsilon}_{pr}(\hat{a}^{\dag}e^{{-i}\omega_{pr}t}-\hat{a}e^{{i}\omega_{pr}t})\nonumber \\
&-{i}{\hbar}{\varepsilon}_{rf}(\hat{b}^{\dag}+\hat{b})(e^{{-i}\omega_{pr}t}-e^{{i}\omega_{pr}t}),
\end{align}
where $g_{om}=\frac {w_a}{L} {\sqrt{\frac{\hbar}{2\omega_b m_b}}}$ is the single-photon coupling strength, and ${\varepsilon}_{rf}=F_{rf}\sqrt{\frac{1}{8\hbar \omega_b m_b}}$ is the amplitude of RF. As a result, the initial two-level energy cavity system becomes a standard three-levelenergy optomechanical system, as shown in Fig. 2. In the frame that rotates at the frequency of the pump field, with respect to $H^{\prime}=\hbar{\omega_{pu}}{{\hat a}^\dag}{\hat a}-\hbar{\omega_{b}}{{\hat b}^\dag}{\hat b}$ and neglecting the fast oscillating terms at the frequencies $\pm 2\omega_a$, $\pm 2\omega_b$, the Hamiltonian of the system is rewritten as
\begin{align}
&H_O=\hbar{\Delta_a}{\hat{a}}^{\dag}{\hat a}+\hbar \omega_b{\hat{b}}^{\dag}{\hat b}-\hbar{g_{om}}\hat{a}^{\dag}\hat{a}({\hat b}^\dag+{\hat b})\nonumber\\
&+{i}{\hbar}{\varepsilon}_{pu}(\hat{a}^{\dag}-\hat{a})+{i}{\hbar}{\varepsilon}_{pr}(\hat{a}^{\dag}e^{-i{\delta}t}-\hat{a}e^{i{\delta}t})+{i}{\hbar}{\varepsilon}_{rf}(\hat{b}^{\dag}-\hat{b}),
\end{align}
where $\Delta_a={\omega}_{a}-{\omega}_{pu}$ is the frequency detuning of the optical cavity from the pump field, and $\delta={\omega_{pr}}-{\omega_{pu}}$ is the frequency detuning of the optical probe field from the pump field.

By combining the amplitude ${q_{0}}$ of the BAR with the creation and annihilation operators, the mean values of the creation and annihilation operators of the BAR are $\langle{{\hat b}}\rangle=\langle{{\hat b}^\dag}\rangle=b_0$ \cite{NVCenterprl2016}, and we can obtain
\begin{eqnarray}
\begin{split}
{q_{0}}=\sqrt{\frac{\hbar}{2\omega_b m_b}}2b_0=\sqrt{\frac{\hbar}{2\omega_b {m_b}}}2\sqrt{n_0},
\end{split}
\end{eqnarray}
 $b_0={q_{0}}\sqrt{\frac{\omega_b {m_b}}{2 \hbar}}$ is the average amplitude of the BAR phonon operator, and $n_0$ is the initial mean phonon number of the BAR produced by the SAW. Moreover, we can obtain the relation of the average BAR phonon amplitude and RF amplitude, which is shown as ${b_{0}}=\frac{{\varepsilon}_{rf}}{2 \omega_b}$.

\section{DYNAMICS OF THE SYSTEM}
In our system, we consider the situation where the intensities of the optical pump field and the probe field satisfy the condition $\varepsilon_{pr} \ll \varepsilon_{pu}$. The system is operated in a resolved sideband regime, which satisfies the condition $\omega_b \sim \kappa_a$ \cite{NatCoherentBalram}. Now, we adopt the quantum Langevin equations (QLEs) for the operators, where the damping and noise terms are supplemented \cite{PRAPhaseJia,PRAEntanglingBarzanjeh}. More importantly, the average amplitude of the BAR phonon operator $b_0$ is supplemented in the quantum Langevin equations. Then, Heisenberg-Langevin equations of the cavity mode and BAR mode can be obtained as
\begin{eqnarray}
\begin{split}
&\dot{\hat a}=-(i\Delta_{a}+\frac{\kappa_a}{2}){\hat a}+ig_{om}{\hat a}({\hat b}^{\dag}+{\hat b})+\varepsilon_{pu}+\varepsilon_{pr}{e^{-i{\delta}t}}+{\hat f},\\
&\dot{\hat b}=-(i\omega_{b}+\frac{\gamma_{b}}{2}){\hat b}+ig_{om}{{\hat a}^\dag}{\hat a}+\varepsilon_{rf}+{\hat \xi}.
\end{split}
\end{eqnarray}
Here, ${\hat f}$ and ${\hat \xi}$ are the quantum and thermal noise operators, respectively \cite{PRAEntanglingBarzanjeh}. $\gamma_{b}$ is the intrinsic damping rate of the BAR.

Then, we linearize the dynamical equations of the operators by assuming $a=a_{s}+\delta{a}$, $b=b_{s}+\delta{b}$, which are both composed of an average amplitude and a fluctuation term. Here, $a_{s}$ and $b_{s}$ are the steady-state values of the operators when only the strong driving field is applied to the system. By assuming $\varepsilon_{pr} \rightarrow0$ and setting all time derivatives to zero, we obtain
\begin{eqnarray}
\begin{split}
&a_{s}=\frac{\varepsilon_{pu}}{i\Delta_{a}^{\prime}+\frac{\kappa_a}{2}},\\
&b_{s}=\frac{ig_{om}{|a_s|}^2+\varepsilon_{rf}}{i\omega_{b}+\frac{\gamma_b}{2}},
\end{split}
\end{eqnarray}
where $\Delta_{a}^{\prime}=\Delta_{a}-g_{om}( b_{s}^{\ast}+b_{s} )$ denotes the effective frequency detuning of the optical pump field from the optical cavity, including the frequency shift caused by the mechanical motion. In our system, to ensure a strong-strength coupling between the optical pump field and the BAR, we consider the situation where the cavity is driven near the red sideband with  $\Delta_{a}^{\prime} \sim \omega_b$ \cite{PRAEITAgarwal}.

Next, by substituting the assumptions $\hat a={a}_{s}+\delta{\hat a}$ and $\hat b={b}_{s}+\delta{\hat b}$ into the nonlinear QLEs and dropping the small nonlinear terms, we can obtain the linearized QLEs, which are
\begin{eqnarray}
\begin{split}
&\dot{\delta{{\hat a}}}=-(i\Delta_{a}^{\prime}+\frac{\kappa_a}{2})\delta{\hat a}+iG_{om}(\delta{b^{\dag}}+\delta{\hat b})+\varepsilon_{p}{e^{-i{\delta}t}}+\hat f,\\
&\dot{\delta{{\hat b}}}=-(i\omega_{b}+\frac{\gamma_{b}}{2})\delta{\hat b}+i(G_{0}^{\ast}{\delta{a}}+G_{0}\delta{a^{\dag}})+\hat \xi,
\end{split}
\end{eqnarray}
where $G_{om}=g_{om}{a_{s}}$ is the total coupling strength between the optical mode and BAR mode.

Furthermore, the fluctuation terms can be rewritten as
\begin{eqnarray}
\begin{split}
\delta{\hat a}=\delta{\hat a_{+}}e^{-i{\delta}t}+\delta{\hat a_{-}}e^{i{\delta}t},\\
\delta{\hat b}=\delta{\hat b_{+}}e^{-i{\delta}t}+\delta{\hat b_{-}}e^{i{\delta}t},\\
\delta{\hat f}=\delta{\hat {f}_{+}}e^{-i{\delta}t}+\delta{\hat {f}_{-}}e^{i{\delta}t},\\
\delta{\hat \xi}=\delta{{\hat \xi}_{+}}e^{-i{\delta}t}+\delta{\hat {\xi}_{-}}e^{i{\delta}t},
\end{split}
\end{eqnarray}
where $\delta{\hat O_{+}}$ and $\delta{\hat O_{-}}$ (with $O=a,b$) correspond to the components at the original frequencies of $\omega_{pr}$ and $2\omega_{pu}-\omega_{pr}$, respectively \cite{PRAEITSumei,PRAPrecisionZhang}. Next we substitute Eq. (13) into Eq. (12) and ignore the second-order small terms by equating coefficients of terms with the same frequency, and the components at the frequencies $\omega_{pr}$ can be obtained as
\begin{eqnarray}
\begin{split}
&\dot {{\delta{\hat a}}}_+=(i\lambda_{a}-\frac{\kappa_a}{2})\delta{\hat a_+}+iG_{om}\delta{\hat b_+}+\varepsilon_{pr}+\delta{\hat f_+},\\
&\dot {\delta{\hat b}}_+=(i\lambda_{b}-\frac{\gamma_{b}}{2})\delta{\hat b_+}+iG_{0}^{\ast}{\delta{\hat a_+}}+\delta{\hat \xi_+},
\end{split}
\end{eqnarray}
where $\lambda_{a}=\delta-\Delta_{a}^{\prime}$, $\lambda_{b}=\delta-\omega_b$.

Next, we take the expectation values of the operators in our system. The noise terms obey the following fluctuations in correlation 
\begin{eqnarray}
\begin{aligned}
\langle{{{{\hat{f}}_{in}(t) \hat{f}}_{in}^{\dag}(t^{\prime})}}\rangle=[N(\omega_{a})+1]\delta(t-t^{\prime}),\\
\langle{{{{\hat{f}}_{in}^{\dag}(t) \hat{f}}_{in}(t^{\prime})}}\rangle=[N(\omega_{a})]\delta(t-t^{\prime}),\\
\langle{{{{\hat{\xi}}_{in}(t) \hat{\xi}}_{in}^{\dag}(t^{\prime})}}\rangle=[N(\omega_{c})+1]\delta(t-t^{\prime}),\\
\langle{{{{\hat{\xi}}_{in}^{\dag}(t) \hat{\xi}}_{in}(t^{\prime})}}\rangle=[N(\omega_{c})]\delta(t-t^{\prime}),\\
\end{aligned}
\end{eqnarray}
where $N(\omega_{a})=[exp(\hbar\omega_{a}/k_{B}T)-1]^{-1}$ is the equilibrium mean thermal photon numbers of the optical fields and $N(\omega_{b})=[exp(\hbar\omega_{b}/k_{B}T)-1]^{-1}$ is the equilibrium mean thermal phonon number of the BAR. 

We can safely assume that the optical field satisfies the condition $\hbar\omega_{a}/k_{B}T\gg1$ at room temperature. For the BAR driven by the SAW, whose frequency is in the GHz regime, the environment temperature in the mK regime, which can be reached inside a dilution refrigerator, is sufficient to ensure that $\hbar\omega_{b}/k_{B}T\gg1$ \cite{RevMooptomechanicsAspelmeyer,Photonsfri}. We assume that our system is operated in the mK temperature regime, and the quantum and thermal noise terms can be ignored. 

Under the mean-field steady-state condition $\langle\dot{\delta{{\hat a}}}\rangle=\langle\dot{\delta{{\hat b}}}\rangle=0$, we can get the solution of $\langle{\delta{\hat a_{+}}}\rangle$, which is
\begin{eqnarray}
\langle{\delta{\hat a_{+}}}\rangle=\frac{(\frac{\gamma_{b}}{2}-i{\lambda_{b}})\varepsilon_{pr}}{(\frac{\kappa_a}{2}-i\lambda_{a})(\frac{\gamma_{b}}{2}-i\lambda_{b})+{G_{om}}^2}.
\end{eqnarray}

Based on the input-output relation, the output field at the probe frequency $\omega_{pr}$ can be expressed as \cite{PRAEITAgarwal,PRAPrecisionZhang}
\begin{eqnarray}
\varepsilon_{out}=\kappa_a \langle\delta{\hat a_{+}}\rangle-\varepsilon_{pr}.
\end{eqnarray}
Then the transmission coefficient $t_{pr}$ of the probe field is given by \cite{ScienceOptomechanicallyWeis,PRAPhaseJia}
\begin{eqnarray}
t_{pr}=\frac{\varepsilon_{out}}{\varepsilon_{pr}}=\frac{\kappa_a \langle\delta{\hat a_{+}}\rangle}{\varepsilon_{pr}}-1.
\end{eqnarray}
By defining $\varepsilon_{T}=\frac{{\kappa_a}{\langle\delta{\hat a_{+}}\rangle}}{\varepsilon_{pr}}$, we can obtain the quadrature $\varepsilon_{T}$ of the output field at frequency $\omega_{pr}$, which is
\begin{eqnarray}
\varepsilon_{T}=\frac{\kappa_a(\frac{\gamma_{b}}{2}-i\lambda_{b})}{(\frac{\kappa_a}{2}-i\lambda_{a})(\frac{\gamma_{b}}{2}-i\lambda_{b})+{G_{om}}^2},
\end{eqnarray}
which has the standard form for the OMIT \cite{PRAEITAgarwal}. The real part $\rm{Re[\varepsilon_T]}$ and imaginary part Im$[\varepsilon_T]$ describe the absorptive and dispersive behaviors of the system, respectively. Correspondingly, the power transmission coefficient is further defined as 
\begin{eqnarray}
T_{pr}={\mid{t_{pr}}\mid}^2.
\end{eqnarray}

Moreover, the phase $\phi_T$ of the output field can be given as \cite{GU2015569}
\begin{eqnarray}
\phi_T=arg[\varepsilon_{T}]=\frac{1}{2i}\rm{Im}(\frac{{\varepsilon_{T}}}{\varepsilon_{T}^{\ast}}).
\end{eqnarray}
In the red sideband region, the rapid phase dispersion can cause the group delay $\tau_T$ of the probe field, which can be expressed as
\begin{eqnarray}
\tau_T=\frac{\partial{\phi_T}}{\partial{\omega_{pr}}}=\rm{Im}[{\frac{1}{\varepsilon_{T}}}{\frac{\partial{\varepsilon_{T}}}{\partial{\omega_{pr}}}}].
\end{eqnarray}

\section{SAW-CONTROLLED OMIT AND PHYSICAL MECHANISM OF THE SYSTEM}

\begin{figure}
\centering
\includegraphics[angle=0,width=9cm]{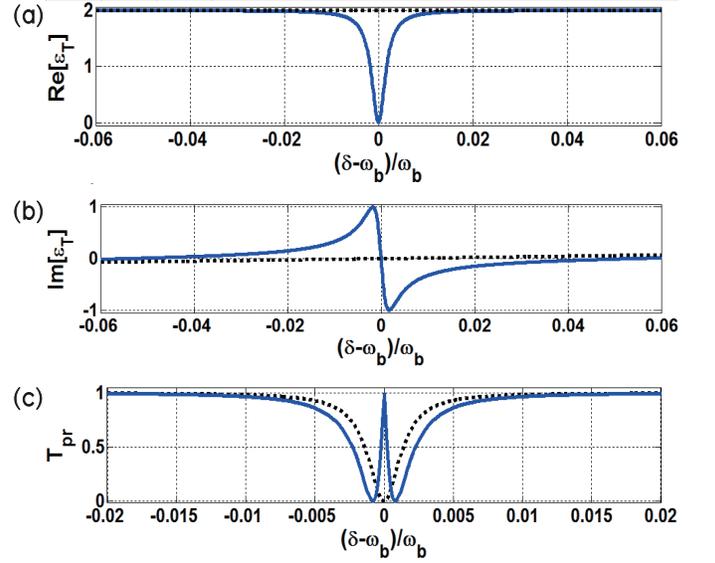}
\caption{(a) Real part Re$[\varepsilon_T]$, (b) imaginary part Im$[\varepsilon_T]$ and (c) power transmission coefficient $T_{pr}$ of the optical probe field as functions of $(\delta-\omega_{b})/\omega_{b}$. The blue-solid line and black-dotted line represent the situations with and without the SAW mode, respectively. The other parameters that we used are $\omega_{a}/{2\pi}=324$ THz, $\kappa_{a}/{2\pi}=3.5$ GHz, $\omega_{b}/{2\pi}=1.05$ GHz, $\gamma_{b} /{2 \pi}= 10.5$ KHz,  $g_{om}/2\pi = 1.54\times 10^7 $ Hz, $P_{pu}=0.015$ $\mu$W, $P_{rf}=0.005$ W.}
\end{figure}

We present below a discussion of the feasibility of the tunable OMIT in the hybrid piezo-optomechanical cavity system. In our system, the parameters that we used are described above, all of which are based on realistic systems. For the high-Q DBR cavity, the thicknesses of the GaAs and AlAs layers are ${\lambda}/{4n_{\rm GaAs}}\sim 64.8$ nm (refractive index $n_{\rm GaAs}=3.57$) and ${\lambda}/{4n_{\rm AlAs}}\sim 77.6$ nm (refractive index $n_{\rm AlAs}=2.98$), respectively. Correspondingly, the $\rm GaAs$ spacer optical thickness of the DBR cavity is $\lambda=925$ nm, the corresponding frequency is $\omega_a = 324$ THz, and the actual thickness is $L = \lambda/n_{\rm GaAs} = 259.1$ nm. To improve the quality of the DBR cavity, referring to the same structure proposed by F. Ding et al., the shape of the defect is defined by Gaussian function $G(x) = \triangle L e^{-x^{2}/{2w^{2}}}$, where $\triangle L$ and $W$ are the thickness and half-width, respectively, as shown in Fig. 1. In our system, we assume that $\triangle L = L/10 = 25.9$ nm and $W = L = 259.1$ nm. We can estimate that the quality factor of our system is increased to $10^5$, and the corresponding cavity linewidth is approximately $3.5$ GHz, so our system satisfies the condition of the sideband-resolved regime $\omega_{b}\sim \kappa_{a}$ \cite{NatCoherentBalram}.

The BAR is driven by the SAW, which is generated by the IDTs, and the length of the IDTs fingers is $l_{\rm{IDTs}}=400\mu m$. The thickness of the upper 10 pairs of Bragg mirrors is approximately 1.42 $\mu$m; as a result, it is within the wavelength scope of the SAW, whose wavelength is $\lambda_s=2.9$ $\mu$m. In our system, the selected frequency of the BAR $\omega_b /{2 \pi}= 1.05$ GHz, which is used in the recent experiment. Based on the related studies, the quality factor of BAR exhibits the order of $10^5$. Accordingly, we can estimate that the intrinsic damping rate of the BAR is $\gamma_{b} /{2 \pi}= 10.5$ kHz, which satisfies the OMIT occurring condition $\kappa_a \gg \gamma_{b}$ \cite{PRAEITAgarwal}. Moreover, their material densities are $\rho_{\rm GaAs}=5.37$ $\rm {g/cm^3}$ and $\rho_{\rm AlAs}=3.72$ $\rm {g/cm^3}$, we can estimate that the average density $\rho_{b} \sim 4.47 \rm {g/cm^3}$ and the effective motional mass of the BAR is ${m_b}=0.33$ pg. Then, based on the parameters that we use, the single-photon coupling strength can be estimated as $g_{om}/2\pi = 1.54\times 10^7 $ Hz.

Furthermore, to ensure that OMIT can be generated, the total optomechanical coupling strength should satisfy the condition $G_{om} \geq {\sqrt{\kappa_{a}\gamma_{b}}/2}$, where a typical transmission window can be obtained \cite{PRAPhaseJia}. Combining this condition with equations (3) and (11) and dropping the small terms, we obtain the maximum power of the RF power, which satisfies the relationship
\begin{eqnarray}
{P_{rf}} \leq (\frac{8 \hbar \pi l_{\rm{IDTs}} v_{\rm{SAW}}^2 \rho_{b}}{m_b})(\varepsilon_{pu}{\sqrt{\frac{1}{\kappa_{a}\gamma_{b}}}}+\frac{\vartriangle_a}{2 g_{om}})^2,
\end{eqnarray}
where the relevant parameters are discussed and provided above. To ensure $b_0 \geq 1$, by combining equations (3) and (9), we obtain the minimum power of the RF power, which satisfies the relationship
\begin{eqnarray}
{P_{rf}} \geq \frac{{8 \hbar {\pi l_{\rm{IDTs}} v_{\rm{SAW}}^2 \rho_{b}}}}{m_b}.
\end{eqnarray}

Next, we discuss the tunable OMIT behaviors of our system. First, based on the selected parameters, Re $[\varepsilon_T]$, Im $[\varepsilon_T]$ and the power transmission coefficient of the optical probe field as a function of $(\delta-\omega_{b})/\omega_{b}$ are plotted in Fig. 3, and the blue-solid line and black-dotted line represent the situations with and without the SAW mode, respectively. When the SAW mode is absent, no transparency window can be obtained in the transmission spectrum curve. However, when the SAW mode is present, a transparency window is obtained in the transmission spectrum curve, whose center position is determined by the frequency point $\delta-\omega_{b}=0$. As a result, our system can be applied in the fields of optical switches, high-resolution spectroscopy and quantum information processing, which have been extensively studied \cite{PRATunableMa,PRAOptomechanicalWang}.

The phenomenon in Fig. 3 arises as follows: when the SAW is absent, the initial system is a two-level cavity system, as shown in Fig. 2(a). It also corresponds to the situation where $g_{om}=0$, referring to equation (3). As a result, the quantum interference between the energy levels cannot induce the occurrence of OMIT, and no transparency window appears. However, when the system is driven by the SAW, the upper Bragg mirrors vibrate as a BAR. Then, the initial two-level cavity system becomes a standard three-level optomechanical system, which is formed by the energy levels of the DBR cavity and BAR, as shown in Fig. 2(b). Under the effect of the optical radiation pressure, the destructive quantum interference between different energy level pathways can be generated. Then, the OMIT phenomenon occurs, and a transparency window is obtained in the transmission spectrum curve. The relevant mechanism has been extensively studied \cite{PRAEITAgarwal,PRATunableMa,PRAOptomechanicalWang}.

\begin{figure}
\centering
\includegraphics[angle=0,width=9cm]{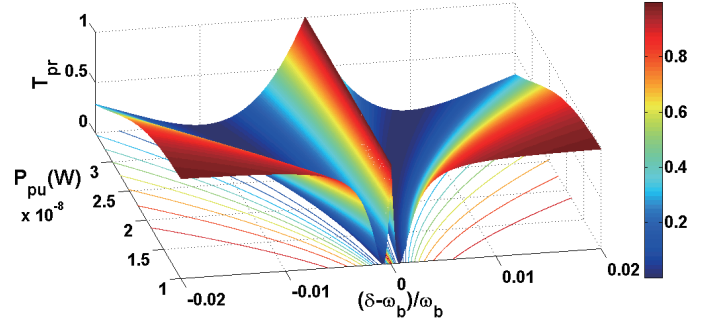}
\caption{(color online) Power transmission coefficient $T_{pr}$ of the optical probe field as functions of $(\delta-\omega_{b})/\omega_{b}$ and pump field power strength $P_{pu}$, $P_{pu}$ is $1 \times 10^{-8} - 3 \times 10^{-8}$ W, and the other parameters are identical to those in Fig. 3.}
\end{figure}

\begin{figure}
\centering
\includegraphics[angle=0,width=9cm]{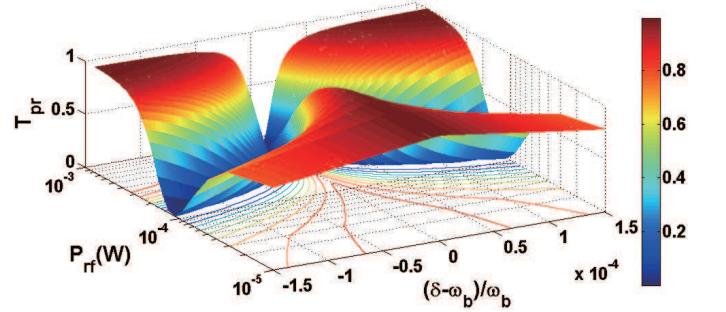}
\caption{(color online) Power transmission coefficient $T_{pr}$ of the optical probe field as functions of $(\delta-\omega_{b})/\omega_{b}$ and RF power $P_{rf}$. $P_{rf}$ is $10^{-5}-10^{-3}$ W, and the other parameters are identical to those in Fig. 3}
\end{figure}

\begin{figure}
\centering
\includegraphics[angle=0,width=9cm]{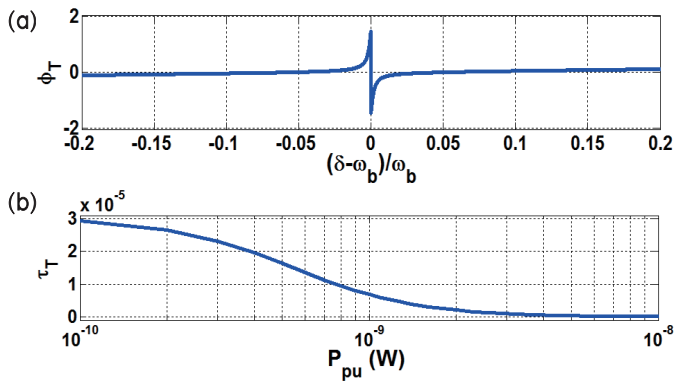}
\caption{ (a) Phase $\phi_T$ of the optical probe field as a function of $(\delta-\omega_{b})/\omega_{b}$. (b) Group delay $\tau_T$ as a function of pump field power strength $P_{pu}$. The other parameters are identical to those in Fig. 3.}
\end{figure}

Fig. 4 presents the power transmission coefficient $T_{pr}$ with respect to $(\delta-\omega_{b})/\omega_{b}$ for different strengths $P_{pu}$ of the optical pump field. In the presence of the SAW, when $P_{pu}$ increases, the transmission window at the frequency position $\delta=\omega_b$ widens. The reason is that in the OMIT situation, the width $\Gamma$ of the transmission window is proportional to the optical pump field power with the relations $\Gamma=\gamma_b+{4{G_{om}}^2}/\kappa_a$ and $G_{om}=g_{om}{a_{s}}$, and the relevant mechanisms have been extensively studied \cite{RevMooptomechanicsAspelmeyer}. As a result, this phenomenon can be applied in the fields of optical switches and optical quantum information process.

Fig. 5 presents the power transmission coefficient $T_{pr}$ with respect to $(\delta-\omega_{b})/\omega_{b}$ for different powers $P_{rf}$ of RF. When the SAW is generated by the IDTs, which is driven by the RF, with the increase in $P_{rf}$, the power transmission coefficient at frequency position $\delta=\omega_b$ decreases. The reason is that in the OMIT situation, the power transmission coefficient $T_{pr}$ is proportional to the total coupling strength $G_{om}=g_{om}{a_{s}}$. Based on equations (3), (9) and (11), the average phonon number $a_{s}$ of the pump field can be determined by $P_{rf}$. With the increase in $P_{rf}$, the total coupling strength is diminished; hence, the power transmission coefficient at frequency position $\delta=\omega_b$ decreases. The minimum and maximum values of the RF power approximately correspond to transmission coefficients of 1 and 0, respectively, as presented in equations (23) and (24). The relevant mechanisms have also been extensively studied \cite{RevMooptomechanicsAspelmeyer}. This phenomenon can also be applied in the field of optical quantum information processing.

To further explore the characteristics of our system, when OMIT occurs, the phase $\phi_{T}$ as functions of $(\delta-\omega_{b})/\omega_b$ and group delay $\tau_T$ as functions of pump field power strength $P_{pu}$ are plotted in Fig. 6. Fig. 6 (a) shows that when OMIT occurs, the phase at frequency position $\delta=\omega_b$ is excessively modulated, which indicates that the group velocity of the probe field is altered. This situation generates a slow-light effect. Fig. 6 (b) shows that with increasing pump field power, the group delay of the probe field decreases. When the power of the pump field is sufficiently weak, the maximal time of the group delay can reach 0.03 ms. As a result, our system can be applied in the field of optical quantum information memory in solid-state systems.

\section{CONCLUSION}

In conclusion, we propose a scheme to generate tunable OMIT in a hybrid piezo-optomechanical cavity system. The system is composed of a high-quality planar DBR cavity modified with an embedded Gaussian-shaped defect. Moreover, IDTs are fabricated on the surface of the cavity and can generate the SAW under RF driving. We show that when a strong pump optical field and a weak probe optical field are applied to the hybrid optomechanical cavity system, in the presence of the SAW, a transmission window can be obtained in the weak output probe field. In the presence of the SAW, the upper Bragg mirrors vibrate as a BAR, and the DBR cavity becomes a lambda-type three-level optomechanical system. As a result, the destructive quantum interference between different energy level pathways generates OMIT, which makes a transmission window appear in the weak output probe field. Conversely, without the SAW, the transmission window in the weak output probe field disappears. Our scheme can be applied in the fields of optical switches, quantum information memory, high-resolution spectroscopy and quantum information processing in solid-state systems.

\begin{acknowledgments}
This work is supported by the Strategic Priority Research Program (Grant No. XDB01010200), the National Natural Sciences Foundation of China (Grants No. 11674337, No. 61605225, and No. 11547035), and the Natural Science Foundation of Shanghai (Grants No. 16ZR1448400), the General Project of National Laboratory Foundation of Radar Signal Processing (Grant No. 61424010106), and the Natural Science Foundation of the Jiangsu Higher Education Institutions
of China (Grant 20KJB140010).
\end{acknowledgments}


\bibliographystyle{apsrev4-1}
\bibliography{doubleOMIT}

\end{document}